\documentclass[prb,aps,showpacs,floatfix,floats,nobalancelastpage,twocolumn]{revtex4}
\usepackage{graphicx}
\usepackage{latexsym,amsmath,amssymb,bm,euscript}
\usepackage{color}
\usepackage{dcolumn}
\usepackage{bm}
\usepackage{epstopdf}
\usepackage{times}
\usepackage{multirow}

\def\ra{\rangle}
\def\la{\langle}
\def\up{\uparrow}
\def\dn{\downarrow}
\def\Hc{{\rm H.c.}}

\def\ET{{$\kappa$-(ET)$_2$Cu$_2$(CN)$_3$}}
\def\dmit{{EtMe$_3$Sb[Pd(dmit)$_2$]$_2$}}
\def\hyperkag{{Na$_4$Ir$_3$O$_8$}}

\begin{document}

\title{Effects of Zeeman field on a Spin Bose-Metal phase}
\author{Hsin-Hua Lai}
\affiliation{Department of Physics, California Institute of Technology, Pasadena, California 91125, USA}
\author{Olexei I. Motrunich}
\affiliation{Department of Physics, California Institute of Technology, Pasadena, California 91125, USA}
\date{\today}

\begin{abstract}
We consider Zeeman field effects on a Spin Bose-Metal (SBM) phase on a two-leg triangular ladder.  This phase was found in a spin-1/2 model with ring exchanges [D. N. Sheng et. al., Phys. Rev. B {\bf 79}, 205112 (2009)], and was also proposed to appear in an interacting electronic model with longer-ranged repulsion [Lai et. al., Phys. Rev. B {\bf 81}, 045105 (2010)].  Using bosonization of a spinon-gauge theory, we study the stability of the SBM phase and its properties under the field.  We also explore phases arising from potential instabilities of the SBM; in all cases, we find a gap to spin-1 excitations while spin-nematic correlations are power law.  We discuss two-dimensional analogues of these phases where spinons can pair with their own species.
\end{abstract}
\maketitle

\section{Introduction}

There has been much recent interest in gapless spin liquids stimulated by several experimental candidates, including two-dimensional (2D) triangular lattice based organic compounds\cite{Shimizu03, Kurosaki05, SYamashita08, MYamashita09, Itou08} \ET\ and \dmit\ and 3D hyper-kagome material\cite{Okamoto07} \hyperkag.  One line of theoretical ideas considers states with a Fermi surface of fermionic spinons.\cite{ringxch, SSLee, Zhou_hypkag, Lawler_hypkag}  For the 2D spin liquids, such a state arises as a good variational wavefunction\cite{ringxch} for a spin model with ring exchanges and is also an appealing candidate for an electronic Hubbard model near the Mott transition.\cite{SSLee, Senthil_Mott, Podolsky09}

Driven by the need for a controlled theoretical access to such phases, Ref.~\onlinecite{Sheng09} considered the Heisenberg plus ring exchanges model\cite{Klironomos, Meyer08} on a two-leg triangular strip -- so-called zigzag chain.  Using numerical  and analytical approaches, Ref.~\onlinecite{Sheng09} found a ladder descendant of the 2D spin liquid in a broad range of parameters and called this phase ``Spin Bose-Metal'' (SBM).  The name refers to metal-like itinerancy present in the spin degrees of freedom (note that there is no electric transport to speak of in the spin-only model).  Further work Ref.~\onlinecite{Lai10} studied electronic Hubbard-type models with longer-ranged repulsion and showed that they are promising systems to realize such an SBM phase in a Mott insulator of electrons proximate to a two-band metallic phase on the zigzag chain.

This paper continues efforts to gain insights about the 2D spin liquid from the solvable 2-leg ladder example.  Here we study the SBM phase under Zeeman magnetic field, while in a separate paper we will study orbital field.  One motivation comes from experiments on the 2D spin liquid materials \ET\ and \dmit\ measuring thermodynamic, transport, and local magnetic properties under strong fields.\cite{Shimizu03, Kawamoto04, Shimizu06, Kawamoto06, SYamashita08, Itou08}  An important question is whether the field can induce changes in the physical state of the system.

To this end, we explore possible instabilities of the 2-leg SBM state in the Zeeman field.  There have been many studies of 2D and 1D spin models under magnetic field showing rich behaviors.  For example, the phase diagram of the $J_1 - J_2$ antiferromagnetic chain with $J_1,~J_2 >0$ in the field\cite{Okunishi99, Okunishi03, McCulloch08, Okunishi08, Hikihara10} contains one-component and two-component Luttinger liquids, a plateau, a phase with static chirality order, and a phase with spin-nematic correlations.  In the spirit of such studies, we allow a large range of fields, which could be numerically explored in spin or electronic models realizing the SBM phase.\cite{Sheng09, Lai10}  We remark that experiments on the spin liquid materials achieve only relatively small fields -- e.g., the maximum magnetization is $\lesssim 0.01 \mu_B$ per spin.  Nevertheless, some of our 2-leg ladder phases from the broader theoretical study motivate interesting 2D states that are worth exploring.

The SBM phase on the zigzag chain can be viewed as a Gutzwiller-projected spinon state where both $\up$ and $\dn$ spinon species populate two Fermi segments (cf.\ Fig.~\ref{dispersion-zeeman}).  The projection eliminates the overall charge mode leaving three gapless modes.  We find that this phase can in principle remain stable under the Zeeman field.  We also identify all possible instabilities out of the SBM.  

Loosely speaking, the instabilities correspond to pairing of spinons separately within each species (a kind of triplet pairing).  More precisely, the relevant interactions can be interpreted as moving a ``Cooper pair'' from one band to the other of the same species.  Of course, there is no long-range pairing order in the quasi-1D and in fact the dominant correlations in our system need not be of ``pair-type'' -- the Bosonization provides the proper treatment, while this language is only for convenience.

It can happen that the pairing is relevant for one spinon species but not the other.  In this case the system retains two gapless modes.  Interestingly, spin-1 excitations become gapped (i.e., transverse spin correlations are short-ranged), while spin-2 excitations are gapless (i.e., nematic or two-magnon correlation functions show power law).

It can also happen that the pairing is relevant for each spinon species.  In this case the system retains only one gapless mode.  Again, spin-1 excitations are gapped while spin-2 remain gapless.  It further turns out that the system breaks translational symmetry and has either period 2 Valence Bond Solid (VBS) or period 2 static chirality order.

Such thinking about pairing within the same spinon species can be extended to 2D.  Here, if we pair only one species and not the other, we have a gap to spin-1 excitations while at the same time we have critical spin-2 correlations and the system retains the gapless Fermi surface for the unpaired species.  On the other hand, if we have pairing within both spinon species, the system acquires a long-range spin-nematic order.\cite{Shindou09}

Spin-nematic phases were discovered and much discussed recently in other interesting frustrated systems.  For instance, such phases were found in the antiferromagnetic zigzag ladder with easy-plane anisotropy\cite{Nersesyan} and in the ferro/antiferro zigzag ladder ($J_1 < 0,~J_2 > 0$) in the Zeeman field.\cite{Chubukov91, Vekua07, Hikihara08, Sato09}  As for examples in 2D, spin-nematic order was found in the frustrated square lattice with ferromagnetic $J_1 < 0$ and antiferromagnetic $J_2 > 0$ and ring exchanges,\cite{Shannon06} and in the triangular lattice with ferromagnetic Heisenberg and antiferromagnetic ring exchanges.\cite{Momoi06} Though, many details of the nematic phases proximate to the SBM studied here are of course different.

The paper is organized as follows.  In Sec.~\ref{weak-coupling:Zeeman}, we consider an electronic Hubbard-type model with longer-ranged repulsion under Zeeman magnetic field and discuss the weak coupling phase diagram in the two-band regime.  We then take a leap to the Mott insulator regime, which can be achieved from the electronic perspective by gapping out the overall charge mode using an eight-fermion Umklapp interaction.  In Secs.~\ref{strong-coupling:Zeeman}-\ref{sec:observable-Zeeman}, we discuss the theory and properties of the SBM under Zeeman field, and in Sec.~\ref{phases nearby SBM: Zeeman} we consider possible instabilities and characterize the resulting phases.  We conclude by discussing generalizations of these phases to 2D.


\section{Electrons on a two-leg zigzag strip in a Zeeman field: Weak coupling approach}\label{weak-coupling:Zeeman}

In this section, we consider half-filled electronic $t_1 - t_2$ chain with extended repulsive interaction in the magnetic Zeeman field.  The Hamiltonian is $H = H_0 + H_Z + H_V$ with
\begin{eqnarray}
\nonumber
&& H_0 = -\sum_{x,\alpha} \big[ t_1 c^\dagger_\alpha(x) c_\alpha(x+1) + t_2 c^\dagger_\alpha(x) c_\alpha(x+2) \\
&& \hspace{6cm} + \Hc \big] ~, \label{freehamiltonian} \\
&& H_Z = -h \sum_x S^z(x) ~, \label{zeemanhamiltonian}\\
&& H_V = \frac{1}{2} \sum_{x,x'} V(x-x') n(x) n(x') ~.
\label{H_V}
\end{eqnarray}
Here $c_\alpha(x)$ is a fermion annihilation operator, $x$ is a site label on the one-dimensional (1D) chain, and $\alpha = \up, \dn$ is a spin index; $n(x) \equiv c^\dagger_\up(x) c_\up(x) + c^\dagger_\dn(x) c_\dn(x)$ is electron number on the site.  Throughout, electrons are at half-filling.  The Zeeman field couples to electron spin $S^z(x) \equiv \frac{1}{2}[c^\dagger_\up(x) c_\up(x) - c^\dagger_\dn(x) c_\dn(x)]$.

\begin{figure}
  \includegraphics[width=\columnwidth]{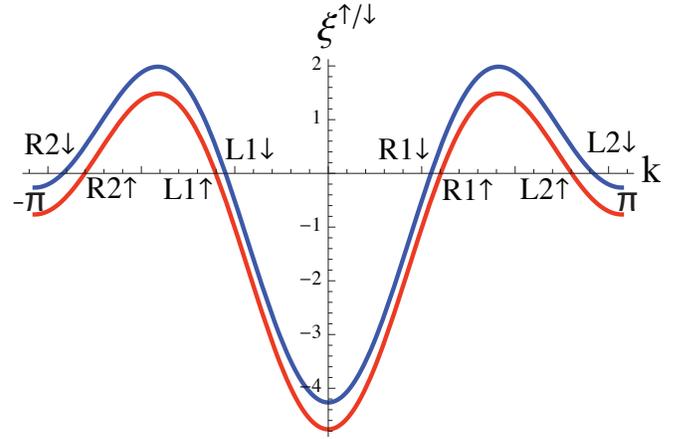}\\
  \caption{
Single-particle spectrum in the presence of the Zeeman field, $\xi^{\up/\dn}(k) = -2 t_1 \cos(k) - 2 t_2 \cos(2k) \mp \frac{h}{2} - \mu$, shown for parameters $t_2/t_1 = 1$ and $h/t_1 = 1/2$.  Our $k_F$-s denote right-moving momenta $\in (-\pi, \pi)$; with this convention, the half-filling condition reads $k_{F1\up} + k_{F1\dn} + k_{F2\up} + k_{F2\dn} = -\pi$.
}
  \label{dispersion-zeeman}
\end{figure}

\begin{figure}[t]
   \includegraphics[width=\columnwidth]{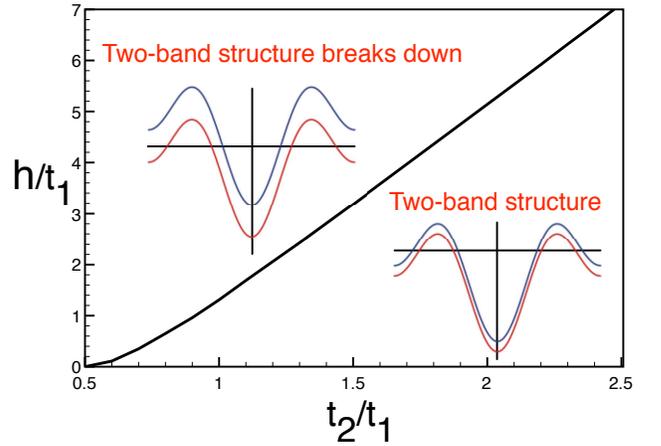}\\
   \caption{
Free electron phase diagram in the $t_2/t_1$--$h/t_1$ plane.  In this paper, we focus solely on the lower region where both spin species have two Fermi seas (``bands'').  For reference, we give the magnetization $M^z \equiv (n_\up - n_\dn)/(n_\up + n_\dn)$ at the transition for several band parameters: $M^z_{\rm crit} = 0.32, 0.46, 0.54$ for $t_2/t_1 = 1.0, 1.5, 2.0$.
}
   \label{t2/t1-h}
\end{figure}

In the weak coupling approach, we assume $H_V \ll H_0, H_Z$ and start with the non-interacting band structure given by $H_0 + H_Z$ and illustrated in Fig.~\ref{dispersion-zeeman}.  In this paper, we focus on the regime $t_2/t_1 > 0.5$ and not too large Zeeman field so that there are two occupied Fermi segments (``bands'') for each spin species.  The corresponding phase boundary in the $t_2/t_1$--$h/t_1$ plane is shown in Fig.~\ref{t2/t1-h}.  For fields exceeding some critical values, the second spin-$\dn$ Fermi segment gets completely depopulated; this regime leads to a different theory and is not considered here.

The spectrum is linearized near the Fermi points and the electron operators are expanded in terms of continuum fields,
\begin{eqnarray}
c_\alpha(x) = \sum_{P,a} e^{i P k_{Fa\alpha} x} c_{Pa\alpha} ~,
\end{eqnarray}
with $P = R/L = +/-$ denoting the right/left movers and $a=1,2$ denoting the two Fermi seas for each spin species, cf.~Fig.~\ref{dispersion-zeeman}.  There are four different Fermi velocities $v_{a\alpha}$.

Using symmetry arguments, we can write down the most general form of the four-fermion interactions which mix the right and left moving fields:
\begin{eqnarray}
H_{\rm int} &=& H_\up + H_\dn + H_{\up\dn} ~, \label{Hint} \\
H_\alpha &=& \lambda_{11}^\alpha \rho_{R1\alpha} \rho_{L1\alpha} + \lambda_{22}^\alpha \rho_{R2\alpha} \rho_{L2\alpha} \\
&+& \lambda_{12}^\alpha (\rho_{R1\alpha} \rho_{L2\alpha} + \rho_{L1\alpha} \rho_{R2\alpha}) \\ 
&+& w_{12}^\alpha (c_{R1\alpha}^\dagger c_{L1\alpha}^\dagger c_{L2\alpha} c_{R2\alpha} + \Hc) ~, \label{w12} \\
H_{\up\dn} &=& \sum_{a, b} \lambda^{\up\dn}_{ab} (\rho_{Ra\up} \rho_{Lb\dn} + \rho_{La\up} \rho_{Rb\dn} )~.
\end{eqnarray}
(Interactions that do not mix right and left movers only shift velocities and do not affect the weak coupling treatment.)

The weak coupling renormalization group (RG) equations are\cite{Muttalib86, Fabrizio1993, Ledermann2000, Meyer08}
\begin{eqnarray}
\dot{\lambda}^\alpha_{11} &=& -\frac{(w^\alpha_{12})^2}{2\pi v_{2\alpha}} ~, \label{lambda11rg}\\
\dot{\lambda}^\alpha_{22} &=& -\frac{(w^\alpha_{12})^2}{2\pi v_{1\alpha}} ~, \label{lambda22rg}\\
\dot{\lambda}^\alpha_{12} &=& \frac{(w^\alpha_{12})^2}{\pi (v_{1\alpha} + v_{2\alpha})} ~, \label{lambda12rg}\\
\dot{w}^\alpha_{12} &=& -\left[ \frac{\lambda^\alpha_{11}}{v_{1\alpha}} + \frac{\lambda^\alpha_{22}}{v_{2\alpha}} - \frac{4 \lambda^\alpha_{12}}{v_{1\alpha} + v_{2\alpha}} \right] \frac{w^\alpha_{12}}{2\pi} ~, \label{w12rg} \\
\dot{\lambda}^{\up\dn}_{ab} &=& 0~.
\end{eqnarray}
Here $\dot{O} \equiv dO/d\ell$, where $\ell$ is logarithm of the length scale; $\alpha = \up,\dn$; and $a,b \in\{1,2\}$.  We see that the terms $\lambda^{\up\dn}_{ab}$ do not flow and the two spin species behave independently from each other in the weak coupling regime.

We therefore focus on one species at a time.  Effectively, this is equivalent to a two-band model of spinless fermions in one dimensions \cite{Muttalib86, Fabrizio1993, Ledermann2000, Meyer08} in the absence of any Umklapps.  The RG Eqs.~(\ref{lambda11rg})-(\ref{w12rg}) have the Kosterlitz-Thouless form and can be solved exactly.  We define
\begin{eqnarray}
y^\alpha \equiv \frac{\lambda^\alpha_{11}}{2\pi v_{1\alpha}} + \frac{\lambda^\alpha_{22}}{2\pi v_{2\alpha}} - \frac{2 \lambda^\alpha_{12}}{\pi (v_{1\alpha} + v_{2\alpha})} ~.
\end{eqnarray}
Eqs.~(\ref{lambda11rg})-(\ref{w12rg}) simplify,
\begin{eqnarray}
\dot{y}^\alpha &=& - \frac{(v_{1\alpha} + v_{2\alpha})^2 + 4 v_{1\alpha} v_{2\alpha}}{2\pi^2 v_{1\alpha} v_{2\alpha}(v_{1\alpha} + v_{2\alpha})^2} \left(w^\alpha_{12} \right)^2 ~,\\
\dot{w}^\alpha_{12} &=& -y^\alpha w^\alpha_{12} ~.
\end{eqnarray}
The $w_{12}^\alpha$ renormalizes to zero if the bare couplings satisfy
\begin{eqnarray}
y^\alpha(\ell=0) \geq \sqrt{\frac{(v_{1\alpha} + v_{2\alpha})^2 + 4 v_{1\alpha} v_{2\alpha}}{2\pi^2 v_{1\alpha} v_{2\alpha} (v_{1\alpha} + v_{2\alpha})^2}} \left|w^\alpha_{12}(\ell=0) \right| ~.\label{stability condition:zeeman}
\end{eqnarray}
In this case, the two-band state of species $\alpha$ is stable and gives two gapless modes.

On the other hand, if the condition Eq.~(\ref{stability condition:zeeman}) is not satisfied, then $w_{12}^\alpha$ runs to strong coupling.  In this case, only one gapless mode remains.  To analyze this, we bosonize
\begin{equation}
c_{Pa\alpha} \sim \eta_{a\alpha} e^{i (\varphi_{a\alpha} + P\theta_{a\alpha})} ~,
\end{equation}
with canonically conjugate boson fields:
\begin{eqnarray}\label{commutation relation}
[\varphi_{a\alpha}(x) , \varphi_{b\beta}(x^\prime)] &=&
[\theta_{a\alpha}(x) , \theta_{b\beta}(x^\prime)] = 0 ~, \\ ~
[\varphi_{a\alpha}(x) , \theta_{b\beta}(x^\prime)] &=& 
i \pi \delta_{ab} \delta_{\alpha\beta} \, \Theta(x - x^\prime) ~,
\end{eqnarray}
where $\Theta(x)$ is the Heaviside step function.  Here we use Majorana fermions ($\{ \eta_{a\alpha}, \eta_{b\beta}\} = 2\delta_{ab} \delta_{\alpha \beta}$) as Klein factors, which assure that the fermion fields with different flavors anti-commute with one another. 

For convenience, we introduce
\begin{eqnarray}
\theta_\alpha^\pm &\equiv& \frac{\theta_{1\alpha} \pm \theta_{2\alpha}}{\sqrt{2}} ~, \quad \alpha=\up \text{or} \dn ~, \label{alpha+-} \\
\theta_{\rho+} &\equiv& \frac{\theta_\up^+ + \theta_\dn^+}{\sqrt{2}} = \frac{\theta_{1\up} + \theta_{2\up} + \theta_{1\dn} + \theta_{2\dn}}{2} ~, \label{rho+} \\
\theta_{\sigma+} &\equiv& \frac{\theta_\up^+ - \theta_\dn^+}{\sqrt{2}} = \frac{\theta_{1\up} + \theta_{2\up} - \theta_{1\dn} - \theta_{2\dn}}{2} ~, \label{sigma+}
\end{eqnarray}
and similarly for $\varphi$ variables.  The $w_{12}^\alpha$ term becomes
\begin{equation}\label{W-terms}
w_{12}^\alpha (c_{R1\alpha}^\dagger c_{L1\alpha}^\dagger c_{L2\alpha} c_{R2\alpha} + \Hc) \sim w_{12}^\alpha \cos(2\sqrt{2} \varphi_\alpha^{-}) ~.
\end{equation}
When $w_{12}^\alpha$ is relevant and flows to large values, it pins the difference field $\varphi_\alpha^-$, while the overall field $\varphi_\alpha^+$ remains gapless (as it should, since the $\alpha$-electrons have an incommensurate conserved density and there are no four-fermion Umklapps).  In this phase, the $\alpha$-electron operator becomes gapped.  Pair-$\alpha$-electron operator is gapless, and also specific particle-hole composites are gapless, with details depending on the sign of $w_{12}^\alpha$.  We are primarily interested in repulsively interacting electrons and expect the particle-hole observables to be more prominent, although not dramatically since for too strong repulsion the conducting state of the $\up$ and $\dn$ electrons is destroyed towards Mott insulator as described below.  We do not provide more detailed characterization of the conducting phases of electrons here, as we are eventually interested in the Mott insulating regime where the $\up$ and $\dn$ species become strongly coupled.  (The two-band spinless electron system was considered, e.g., in Refs.~\onlinecite{Muttalib86, Fabrizio1993, Ledermann2000, Meyer08}, and our analysis in Sec.~\ref{sec:observable-Zeeman} can be readily tailored to the electronic phases here.)

In the model with longer-ranged density-density repulsion, Eq.~(\ref{H_V}), the bare couplings are
\begin{eqnarray}
\lambda^\alpha_{11} &=& V_{Q=0} - V_{2k_{F1\alpha}} ~,\\
\lambda^\alpha_{22} &=& V_{Q=0} - V_{2k_{F2\alpha}} ~,\\
\lambda^\alpha_{12} &=& V_{Q=0} - V_{k_{F1\alpha} + k_{F2\alpha}} ~,\\
w^\alpha_{12} &=& V_{k_{F1\alpha} - k_{F2\alpha}} - V_{k_{F1\alpha} + k_{F2\alpha}} ~,\\
\lambda^{\up\dn}_{ab} &=& V_{Q=0} ~.
\end{eqnarray}
Here $V_Q \equiv \sum_{x' = -\infty}^\infty V(x-x') e^{i Q (x-x')} = V_{-Q}$.

As an example, we consider the following potential
\begin{eqnarray}
\label{parameterization}
V(x-x') = \left\{
\begin{array}{cc}
U &, \hspace{0.5cm} |x-x'|=0 \\
\kappa U e^{-\gamma |x-x'|} &, \hspace{0.5cm} |x-x'| \geq 1
\end{array}
\right\}
\end{eqnarray}
This was used in Ref.~\onlinecite{Lai10} to provide stable realizations of the C2S2 metal and the SBM Mott insulator of electrons in zero field.  Here $U$ is the overall energy scale and also the on-site repulsion; dimensionless parameter $\kappa$ controls the relative strength of further-neighbor interactions; and $\gamma$ defines the decay rate.  Applying the stability condition, Eq.~(\ref{stability condition:zeeman}), we can now determine the phase diagram in the weak coupling approach in the regime where the kinetic energy gives four modes.

\begin{figure}[t]
  \includegraphics[width=\columnwidth]{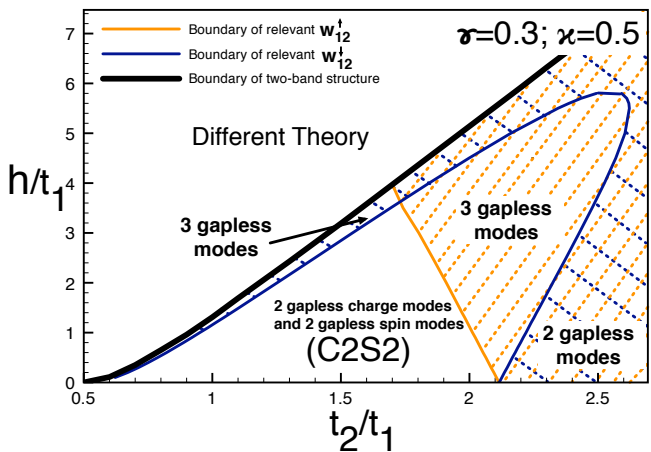}\\
  \caption{
An example of the weak-coupling phase diagram in the electron system under the Zeeman field, using model interactions Eq.~(\ref{parameterization}) with $\kappa = 0.5$ and $\gamma = 0.3$.  We focus on the region where the kinetic energy gives four modes (cf.\ Figs.~\ref{dispersion-zeeman},~\ref{t2/t1-h}) and find four phases: metallic phase with four gapless modes evolving out of the C2S2 phase in zero field; phase with three gapless modes where only the $w^\up_{12}$-term is relevant and flows to strong coupling; phase with three gapless modes where only the $w^\dn_{12}$-term is relevant; and phase with two gapless modes where both the $w^\up_{12}$ and $w^\dn_{12}$ are relevant.  The $w^\up_{12}$-term is relevant in the region with hash lines at roughly 45 degrees and the $w^\dn_{12}$-term is relevant in the region with hash lines at 135 degrees with respect to the horizontal axis.  Note that the $w^\dn_{12}$-term always becomes relevant upon approaching the boundary of the two-band structure.\cite{Meyer08}  
}
  \label{weak coupling phase diagram:zeeman}
\end{figure}

Figure~\ref{weak coupling phase diagram:zeeman} provides an illustration for $\gamma = 0.3$ and $\kappa = 0.5$.  The $w_{12}^\up$ interaction is relevant in the region with hash lines at roughly 45 degrees and the $w_{12}^\dn$ is relevant in the region with hash lines at 135 degrees with respect to the horizontal axis.  There are four distinct phases.  First, when both $w_{12}^\up$ and $w_{12}^\dn$ are irrelevant, we have a phase with four gapless modes, which is connected to the C2S2 phase at $h=0$.  (Note, however, that we assumed $H_Z \gg H_V$, so the formal $h \to 0$ limit here is different from the weak coupling analysis at $h=0$ in Refs.~\onlinecite{Louis01, Lai10}.)

Next, when $w_{12}^\up$ is relevant while $w_{12}^\dn$ is irrelevant, we have a phase with three gapless modes: one associated with the $\up$-electrons and two associated with the $\dn$-electrons.  In this phase, inserting a single $\up$-electron costs a finite gap while inserting a pair of $\up$-electrons or a particle-hole combination of $\up$-electrons is gapless.  The $\dn$ electrons are completely gapless.  

When $w_{12}^\dn$ is relevant while $w_{12}^\up$ is irrelevant, we have another phase with three gapless modes, which is similar to the preceding paragraph but with $\up$ and $\dn$ interchanged.  As can be seen in Fig.~\ref{weak coupling phase diagram:zeeman}, $w_{12}^\dn$ is always relevant when $h$ approaches the critical value,\cite{Meyer08} and the instability arises because the $v_{2\dn}$ approaches zero.

Finally, for large $t_2/t_1$, both $w_{12}^\up$ and $w_{12}^\dn$ are relevant and we have a phase with only two gapless modes: one associated with spin-$\up$ and the other with spin-$\dn$ species.  In this case, inserting a single electron of either spin is gapped, while inserting a pair or a particle-hole combination of same-spin electrons is gapless.

\section{Transition to Mott Insulator: SBM phase}
\label{strong-coupling:Zeeman}

Note that all phases accessed from the weak coupling analysis are conducting along the zigzag chain.  Mott insulating states do not appear since there is no four-fermion Umklapp.  The half-filled system does become insulating for sufficiently strong repulsion.  This can be achieved by including a valid eight-fermion Umklapp, which is irrelevant at weak coupling but can become relevant at intermediate to strong coupling:\cite{Sheng09, Lai10}
\begin{eqnarray}
H_8 &=& v_8
(c_{R1\up}^\dagger c_{R1\dn}^\dagger c_{R2\up}^\dagger c_{R2\dn}^\dagger
 c_{L1\up} c_{L1\dn} c_{L2\up} c_{L2\dn} + \Hc )
\nonumber \\
 &\sim& 2 v_8 \cos(4 \theta_{\rho+}) ~,\label{H8bosonized}
\end{eqnarray}
where $\theta_{\rho +}$ is defined in Eq.~(\ref{rho+}) and describes slowly varying electron density, $\rho_e(x) = 2 \partial_x \theta_{\rho +}/\pi$.  The density-density repulsion gives coarse-grained interaction $H_{\rm int} \sim V_{Q=0} (\partial_x \theta_{\rho +})^2$.  This will stiffen the $\theta_{\rho +}$ field and will reduce the scaling dimension of the Umklapp term.  For sufficiently strong repulsion the Umklapp becomes relevant and will grow at long scales, pinning the $\theta_{\rho+}$ and driving a metal-insulator transition.  As discussed in Refs.~\onlinecite{Sheng09, Lai10}, we expect that Mott insulator corresponding to a spin model with spins residing on sites is described by $v_8 > 0$ and the pinning condition
\begin{eqnarray}
4\theta_{\rho+}^{(0)} = \pi~~({\rm mod}~2\pi) ~. \label{u8pos}
\end{eqnarray}

Such gapping out of the overall charge mode can occur out of any of the four conducting phases discussed in Fig.~\ref{weak coupling phase diagram:zeeman}.  When this happens out of the four-mode metal, we obtain spin liquid Mott insulator with three gapless modes -- the Spin Bose-Metal.  In principle, one could perform an intermediate coupling analysis similar to that in Ref.~\onlinecite{Lai10} to estimate the strength of the repulsion needed to drive the metal-insulator transition, but we will not try this here.  Below we discuss qualitatively the stability and physical observables in the SBM phase under the Zeeman field.  We will then consider instabilities of the SBM similar to the $w_{12}^\alpha$-driven transitions out of the four-mode metal above, but now with the $\up$ and $\dn$ systems strongly coupled to form the Mott insulator.  

Reference~\onlinecite{Sheng09} also presented another route to describe the SBM in a spin-only model by using Bosonization to analyze slave particle gauge theory.  The formalism is similar to the electron model analysis, but with electron operators $c_\alpha(x)$ replaced with spinon operators $f_\alpha(x)$ and the gauge theory constraint realized via an explicit mass term for $\theta_{\rho+}$,
\begin{eqnarray}
\mathcal{L}_{\rm gauge\ theory} = m \left(\theta_{\rho+} - \theta_{\rho+}^{(0)}\right)^2 ~.
\label{LGT}
\end{eqnarray}
Loosely speaking, spinons are electrons that shed their overall charge once the Umklapp term $H_8$ became relevant.  Note,\cite{Sheng09} however, that in the spin-only model, there are no free spinons, unlike the situation in the electronic model where we have electron excitations above the charge gap. 

From now on, we will use the spinon-gauge language.  To get some quantitative example, we consider the case where spinons do not have any interactions other than Eq.~(\ref{LGT}), i.e., all residual interactions like Eq.~(\ref{Hint}) are set to zero.
Once the $\theta_{\rho+}$ field is pinned and after integrating out the $\varphi_{\rho+}$, we obtain an effective action for the remaining fields $(\theta_{\sigma+},~\theta^{-}_\up,~\theta^{-}_\dn) \equiv \bm{\Theta}^T$ and $(\varphi_{\sigma+},~\varphi^{-}_\up,~\varphi^{-}_\dn)\equiv \bm{\Phi}^T$ defined in Eqs.~(\ref{alpha+-})-(\ref{sigma+}):
\begin{eqnarray}\label{effectL}
\mathcal{L}_{\rm eff} &=& \frac{1}{2\pi} \bigg{[} \partial_x \bm{\Theta}^T \cdot {\bf A} \cdot \partial_x \bm{\Theta} + \partial_x \bm{\Phi}^T \cdot {\bf B} \cdot \partial_x \bm{\Phi}\bigg{]} \\
&& +\frac{i}{\pi} \partial_x \bm{\Theta}^T \cdot \partial_\tau \bm{\Phi} ~.
\end{eqnarray}
Matrix elements of ${\bf A}$ and ${\bf B}$ are,
\begin{eqnarray*}
&& \hspace{-0.9cm} {\bf A} =
\begin{pmatrix}
\bar{v} & \frac{v^{-}_{\uparrow}}{\sqrt{2}}& -\frac{v^{-}_{\downarrow}}{\sqrt{2}}\\
\frac{v^{-}_{\uparrow}}{\sqrt{2}} & v^{+}_{\uparrow} & 0\\
-\frac{v^{-}_{\downarrow}}{\sqrt{2}} & 0 & v^{+}_{\downarrow}
\end{pmatrix}~, \\
&& \hspace{-0.9cm}{\bf B} =
\begin{pmatrix}
\frac{v^{+}_{\uparrow} v^{+}_{\downarrow}}{\bar{v}} & \frac{v^{+}_{\downarrow} v^{-}_{\uparrow}}{\sqrt{2}\bar{v}} & -\frac{v^{+}_{\uparrow} v^{-}_{\downarrow}}{\sqrt{2}\bar{v}}\\
\frac{v^{+}_{\downarrow} v^{-}_{\uparrow}}{\sqrt{2}\bar{v}} & \frac{(v^{+}_{\uparrow})^2 - (v^{-}_{\uparrow})^2+v^{+}_{\uparrow}v^{+}_{\downarrow}}{2\bar{v}} & -\frac{v^{-}_{\uparrow} v^{-}_{\downarrow}}{2\bar{v}} \\
-\frac{v^{+}_{\uparrow} v^{-}_{\downarrow}}{\sqrt{2}\bar{v}} & -\frac{v^{-}_{\uparrow} v^{-}_{\downarrow}}{2\bar{v}} & \frac{(v^{+}_{\downarrow})^2 - (v^{-}_{\downarrow})^2 + v^{+}_{\uparrow} v^{+}_{\downarrow}}{2\bar{v}}
\end{pmatrix},
\end{eqnarray*}
where
\begin{eqnarray}
v^\pm_\alpha &\equiv& \frac{v_{1\alpha} \pm v_{2\alpha}}{2}~, \quad \alpha=\up, \dn;\\
\bar{v} &\equiv& \frac{v^{+}_{\uparrow} + v^{+}_{\downarrow}}{2} = \frac{v_{1\uparrow} + v_{2\uparrow} + v_{1\downarrow} + v_{2\downarrow}}{4}.
\end{eqnarray}
Having all the matrix elements, we can numerically calculate the scaling dimensions of the $w_{12}^\alpha$-terms in Eq.~(\ref{W-terms}).

\begin{figure}[t]
   \includegraphics[width=\columnwidth]{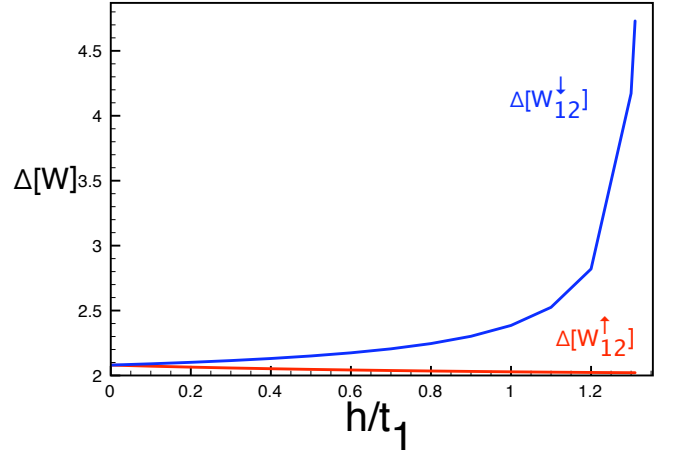}\\
   \caption{
Scaling dimensions $\Delta[w_{12}^\up]$ and $\Delta[w_{12}^\dn]$ as a function of $h/t_1$ for fixed $t_2/t_1 = 1$, calculated in the absence of residual spinon interactions.  In this case, the scaling dimensions stay greater than $2$ and the SBM phase remains stable under the Zeeman field.
}
   \label{ScalW-h1.0}
\end{figure}

As an illustration, Fig.~\ref{ScalW-h1.0} shows the results along a vertical cut at $t_2/t_1 = 1$ from Fig.~\ref{weak coupling phase diagram:zeeman} (assumed driven into the Mott insulator as described above).  We see that in the absence of the residual interactions the SBM remains stable under the Zeeman field.  We also note that the scaling dimensions of the $w^\up_{12}$ and $w^\dn_{12}$ have opposite trends, which implies that the overall stability is reduced.  Since the scaling dimension of the $w^\up_{12}$ interaction decreases with increasing field, it is likely that this will be the first instability channel upon including the residual interactions.  This finding is similar to the weak coupling analysis where the $\up$-system tends to become unstable first.  We want to emphasize, however, that neglecting the residual spinon interactions is likely a poor approximation for any realistic spin model, and any calculations in this scheme should be taken with caution.  The only precise statement here is that the SBM can in principle remain stable under the Zeeman field.

In Sec.~\ref{phases nearby SBM: Zeeman} we discuss phases proximate to the SBM.  Motivated by the above observations, we will consider first the case where only the $w^\up_{12}$ term becomes relevant; we will also consider the situation where both $w^\up_{12}$ and $w^\dn_{12}$ are relevant.  Before this, we need to describe main physical observables in the SBM under the Zeeman field, which we will then use to analyze the instabilities and the properties of the resulting phases.

\section{Observables in the SBM in Zeeman field}
\label{sec:observable-Zeeman}

In the presence of the Zeeman field, the system has $S^z_{\rm tot}$ spin conservation symmetry and complex conjugation symmetry (${\cal C}: i \to -i$) in the $S^z$ basis.  The system also has lattice translation and inversion ($I: x \to -x$) symmetries.  The internal symmetries are sufficiently reduced compared with the SU(2)-invariant case of Ref.~\onlinecite{Sheng09} that we need to revisit the physical observables in the SBM.

We first consider $S^z$-conserving bilinears, which we will also call ``spin-0'' objects,
\begin{eqnarray}
  \epsilon_{2k_{Fa\alpha}} &\equiv& f^\dagger_{La\alpha} f_{Ra\alpha} ~, \\
  \epsilon_{k_{F1\alpha} + k_{F2\alpha}} &\equiv& \frac{1}{2}\left( f^\dagger_{L1\alpha} f_{R2\alpha} + f^\dagger_{L2\alpha} f_{R1\alpha}\right) ~, \\
  \chi_{k_{F1\alpha} + k_{F2\alpha}} &\equiv& \frac{1}{2} \left( f^\dagger_{L1\alpha} f_{R2\alpha} - f^\dagger_{L2\alpha} f_{R1\alpha} \right)~, \\
  \epsilon_{k_{F1\alpha} - k_{F2\alpha}} &\equiv& \frac{1}{2} \left( f^\dagger_{L1\alpha} f_{L2\alpha} + f^\dagger_{R2\alpha} f_{R1\alpha} \right)~, \\
  \chi_{k_{F1\alpha} - k_{F2\alpha}} &\equiv& \frac{1}{2} \left( f^\dagger_{L1\alpha} f_{L2\alpha} - f^\dagger_{R2\alpha} f_{R1\alpha} \right)~,
\end{eqnarray}
(no summation over $a$ or $\alpha$).  We define $\epsilon_{-Q} = \epsilon_Q^\dagger$ and $\chi_{-Q} = \chi_Q^\dagger$ so that $\epsilon(x)$ and $\chi(x)$ are Hermitian operators.

The $\epsilon$ bilinears appear, e.g., when expressing spinon hopping energies, while the $\chi$ bilinears appear in currents.  Specifically, consider a bond $[x, x+n]$,
\begin{eqnarray}
  {\cal B}^{(n)}(x) &\sim& f^\dagger_\alpha(x) f_\alpha(x+n) + \Hc ~, \\
  {\cal J}^{(n)}(x) &\sim& i \left[f^\dagger_\alpha(x) f_\alpha(x+n) - \Hc \right] ~,
\end{eqnarray}
where $\alpha=\up$ or $\dn$ species can come with independent amplitudes.  Expansion in terms of the continuum fields gives, up to real factors,
\begin{eqnarray}
  {\cal B}^{(n)}_Q \sim e^{inQ/2} \epsilon_Q ~, \label{BnQ} \\
  {\cal J}^{(n)}_Q \sim e^{inQ/2} \chi_Q ~. \label{JnQ}
\end{eqnarray}

Note that we can view $\epsilon(x)$ as a site-centered energy operator, e.g., $\epsilon(x) \sim {\cal B}^{(1)}(x-1) + {\cal B}^{(1)}(x) \sim {\cal B}^{(2)}(x-1)$, in the sense of having the same symmetry properties.  We can also view $\epsilon(x) \sim S^z(x)$ in the same sense because of the presence of the Zeeman energy.  [More generally, the symmetry properties of any operator are not changed upon multiplying by $S^z(x)$.]  On the other hand, the bond operator ${\cal B}^{(n)}(x)$ has the same symmetry properties as a bond energy such as $\vec{S}(x) \cdot \vec{S}(x+n)$ and can be used to characterize VBS correlations in the spin system.

Similarly, we can view $\chi(x)$ as a site-centered current, $\chi(x) \sim {\cal J}^{(1)}(x-1) + {\cal J}^{(1)}(x) \sim {\cal J}^{(2)}(x-1)$, and also as a scalar chirality, $\chi(x) \sim \vec{S}(x-1) \cdot \vec{S}(x) \times \vec{S}(x+1)$, while ${\cal J}^{(n)}(x)$ has the same symmetry properties as a spin current, ${\cal J}^{(n)}(x) \sim i [S^+(x) S^-(x+n) - \Hc]$.

Symmetry analysis shows that $\epsilon_Q$ transforms to $\epsilon_{-Q}$ under either lattice inversion $I$ or complex conjugation ${\cal C}$, while $\chi_Q$ transforms to $-\chi_{-Q}$ under either $I$ or ${\cal C}$.  We can then give an independent argument for the relations~Eqs.~(\ref{BnQ}) and (\ref{JnQ}) for $Q \neq 0, \pi$, and can show generally that, up to complex phase factors, such $\epsilon_Q$ and $\chi_Q$ cover all independent spin-0 observables for the system in the Zeeman field.

Special care is needed for $Q=\pi$.  In this case, Eqs.~(\ref{BnQ}) and (\ref{JnQ}) hold only for $n={\rm even}$.  On the other hand, ${\cal B}^{(n={\rm odd})}_\pi$ is odd under inversion $I$ and even under complex conjugation ${\cal C}$, while ${\cal J}^{(n={\rm odd})}_\pi$ is even under $I$ and odd under ${\cal C}$.  In particular, the nearest-neighbor bond ${\cal B}^{(1)}_\pi$ and ${\cal J}^{(1)}_\pi$ are independent observables from $\epsilon_\pi \sim {\cal B}^{(2)}_\pi$ and $\chi_\pi \sim {\cal J}^{(2)}_\pi$.  In the present SBM problem, such $Q=\pi$ observables do not appear as bilinears but appear as four-fermion terms below.

The bosonized expressions for the spin-0 bilinears are:
\begin{eqnarray}
\epsilon_{2k_{Fa\alpha}} &=& i e^{i (\theta_{\rho+} + \alpha \theta_{\sigma+} + a \sqrt{2}\theta_\alpha^-)} ~, \label{epschi_first}\\
\epsilon_{k_{F1\alpha} + k_{F2\alpha}} &=& -i \eta_{1\alpha} \eta_{2\alpha} e^{i (\theta_{\rho+} + \alpha \theta_{\sigma+})} \sin(\sqrt{2}\varphi_\alpha^-) ~, \label{eps1+2}\\
\chi_{k_{F1\alpha} + k_{F2\alpha}} &=& \eta_{1\alpha} \eta_{2\alpha} e^{i (\theta_{\rho+} + \alpha \theta_{\sigma+})} \cos(\sqrt{2}\varphi_\alpha^-) ~,\label{chi1+2} \\
\epsilon_{k_{F1\alpha} - k_{F2\alpha}} &=& -i \eta_{1\alpha} \eta_{2\alpha} e^{i \sqrt{2} \theta_\alpha^-} \sin(\sqrt{2} \varphi_\alpha^-) ~, \\
\chi_{k_{F1\alpha} - k_{F2\alpha}} &=& \eta_{1\alpha} \eta_{2\alpha} e^{i \sqrt{2} \theta_\alpha^-} \cos(\sqrt{2} \varphi_\alpha^-) ~, \label{epschi_last}
\end{eqnarray}
where we used definitions Eqs.~(\ref{alpha+-})-(\ref{sigma+}) and $\alpha = +/-$ for spin $\up$ or $\dn$ and $a = +/-$ for band $1$ or $2$.

To bring out the wavevector $Q=\pi$ that will play an important role in the analysis of phases near the SBM, we need to consider four-fermion terms.  We find,
\begin{eqnarray}
{\cal B}^{(1)}_\pi: && i(\epsilon_{k_{F1\up} + k_{F2\up}} \epsilon_{k_{F1\dn} + k_{F2\dn}} - \Hc) \sim \label{B1pi1}\\
&& \sim \hat{\Gamma} \sin(\sqrt{2}\varphi_\up^-) \sin(\sqrt{2}\varphi_\dn^-) \sin(2\theta_{\rho+}); \\
&& i(\chi_{k_{F1\up} + k_{F2\up}} \chi_{k_{F1\dn} + k_{F2\dn}} - \Hc) \sim \label{B1pi2} \\
&& \sim \hat{\Gamma} \cos(\sqrt{2}\varphi_\up^-) \cos(\sqrt{2}\varphi_\dn^-) \sin(2\theta_{\rho+});
\end{eqnarray}
and also
\begin{eqnarray}
\chi_\pi: && \epsilon_{k_{F1\up} + k_{F2\up}} \chi_{k_{F1\dn} + k_{F2\dn}} + \Hc \sim \label{chipi1}\\
&& \sim \hat{\Gamma} \sin(\sqrt{2}\varphi_\up^-) \cos(\sqrt{2}\varphi_\dn^-) \sin(2\theta_{\rho+}); \\
 && \chi_{k_{F1\up} + k_{F2\up}} \epsilon_{k_{F1\dn} + k_{F2\dn}} + \Hc \sim \label{chipi2} \\
&& \sim \hat{\Gamma} \cos(\sqrt{2}\varphi_\up^-) \sin(\sqrt{2}\varphi_\dn^-) \sin(2\theta_{\rho+}) ~. \label{chipi2b}
\end{eqnarray}
Here $\hat{\Gamma} \equiv \eta_{1\up} \eta_{1\dn} \eta_{2\up} \eta_{2\dn}$.  Note that we have only listed observables containing $\sin(2\theta_{\rho+})$.  The other independent spin-0 objects $\epsilon_\pi$ and ${\cal J}^{(1)}_\pi$ contain $\cos(2\theta_{\rho+})$ and vanish because of the pinning condition Eq.~(\ref{u8pos}).

Having discussed $S^z$-conserving observables, we can similarly consider $S^z$-raising observables.  We will call objects corresponding to $\delta S^z = 1$ or $2$ as ``spin-1'' or ``spin-2'' respectively.  We have spin-1 bilinears,
\begin{equation}
S^+_{-Pk_{Fa\up} + P'k_{Fb\dn}} \equiv f_{Pa\up}^\dagger f_{P'b\dn} ~.
\end{equation}
Generically, these all carry different momenta. We can readily write bosonized expressions.  For reference, we give the main ones that contain oppositely moving fields:
\begin{eqnarray}
S^+_{k_{Fa\up} + k_{Fb\dn}} = &\eta_{a\up} \eta_{b\dn}& e^{-i [\varphi_{\sigma+} + \frac{1}{\sqrt{2}} (a \varphi_\up^- - b \varphi_\dn^-)]} \nonumber \\
&\times& e^{i [\theta_{\rho+} + \frac{1}{\sqrt{2}} (a \theta_\up^- + b \theta_\dn^-)]} ~,
\end{eqnarray}
where we used convention $a,b = +/-$ for band $1$ or $2$.  We can generally argue that at $Q \neq 0,\pi$, objects $S_Q^+$ that transform like Fourier modes of the $S^+(x)$ operator cover, up to complex phases, all distinct spin-1 observables.  In the present SBM system, we do not find any interesting spin-1 observables at $Q=0,\pi$. 

Since we will encounter phases where $S^+$ is gapped, we also need to consider $\delta S^z = 2$ observables, i.e., some kind of ``magnon pair'' creation operators.  Because of the hard spin condition, we define them on bonds $[x,x+n]$,
\begin{eqnarray}
{\cal P}^{+, (n)}(x) &\equiv& S^+(x) S^+(x+n) \\
&\sim& f_\up^\dagger(x) f_\up^\dagger(x+n) f_\dn(x+n) f_\dn(x) ~.
\end{eqnarray}
The last line can be expanded in terms of the continuum fields and organized as follows.  For $\alpha$-species, a ``pair'' operator $f_\alpha^\dagger(x) f_\alpha^\dagger(x+n)$ contains zero momentum contributions $f_{Ra\alpha}^\dagger f_{La\alpha}^\dagger$, $a=1$ or $2$; $\pm (k_{F1\alpha} + k_{F2\alpha})$ momentum contributions $f_{P1\alpha}^\dagger f_{P2\alpha}^\dagger$, $P=L/R$; and $\pm (k_{F1\alpha} - k_{F2\alpha})$ contributions $f_{P1\alpha}^\dagger f_{-P2\alpha}^\dagger$.  Multiplying the pair creation operator for $\up$ species and pair destruction operator for $\dn$ species, we obtain contributions to ${\cal P}^+$ carrying combinations of the above momenta.

We can argue on general symmetry grounds that, up to complex phases, there is a single independent spin-2 object at $Q \neq 0,\pi$.  On the other hand, at $Q=\pi$ there are two independent objects that transform differently under lattice inversion; they can be realized by ${\cal P}^{+,(n={\rm even})}_\pi$ and ${\cal P}^{+,(n={\rm odd})}_\pi$ respectively.  At $Q=0$, we consider only objects ${\cal P}^{+,(n)}_{Q=0}$ which have the same symmetry properties for any $n$.

In the present SBM problem, the main spin-2 observables occur precisely at $Q = 0, \pi$, and we give bosonized expressions only for these.  For $Q=0$, there are four possible terms: 
\begin{eqnarray}
{\cal P}^{+,(n)}_{Q=0} : f^\dagger_{Ra\up} f^\dagger_{La\up} f_{Lb\dn} f_{Rb\dn} \sim e^{-i [2\varphi_{\sigma+} + \sqrt{2}(a \varphi_\up^- - b \varphi_\dn^-)]}
\end{eqnarray}
with independent $a,b = +/-$ corresponding to bands $1$ or $2$.
For $Q=\pi$ we find
\begin{eqnarray}
\nonumber {\cal P}^{+,(n)}_{Q=\pi} \!\!\!&\sim&\!\!\!
f^\dagger_{R1\up} f^\dagger_{R2\up} f_{L2\dn} f_{L1\dn} e^{i\frac{\pi n}{2}} \!+\! f^\dagger_{L1\up} f^\dagger_{L2\up} f_{R2\dn} f_{R1\dn} e^{-i\frac{\pi n}{2}} ~\\
&\sim&\! \hat{\Gamma}\; e^{-i 2\varphi_{\sigma+}} \sin\left[2\theta_{\rho+} + \frac{\pi}{2}(n-1)\right] ~.
\end{eqnarray}
Because of the pinning condition on the $\theta_{\rho+}$, only the ${\cal P}^{+,(n={\rm odd})}_{Q=\pi}$ are non-zero, and we can use the nearest-neighbor magnon pair operator ${\cal P}^{+,(1)}$ as the main representative.

\section{Nearby phases out of the SBM in the field} \label{phases nearby SBM: Zeeman}

We now consider what happens when either $w^\up_{12}$ or $w^\dn_{12}$ from Eq.~(\ref{W-terms}) or both become relevant. 

\subsection{Phases when $w^\up_{12}$ is relevant}
\label{onew12relevant}

Let us start with the case when the $w^\up_{12}$ term is relevant while $w^\dn_{12}$ is irrelevant.  The field $\varphi_\up^-$ is pinned, while fields $\varphi_\dn^-$ and $\varphi_{\sigma+}$ remain gapless, so we have two gapless modes.  There is no static order.  We summarize characteristic power law observables in Table~\ref{tab:up-relevant-comm} and discuss them in turn.

First, all observables $\epsilon_Q$ and $\chi_Q$ in Eqs.~(\ref{epschi_first})-(\ref{epschi_last}) constructed out of the $f_\dn$ fields show power law.  On the other hand, such observables constructed out of the $f_\up$ fields that contain $\theta_\up^-$ become short-ranged once we pin the conjugate $\varphi_\up^-$; thus, only $Q = k_{F1\up} + k_{F2\up}$ can remain power law.  There are two cases depending on the sign of $w^\up_{12}$:
\begin{eqnarray}
w_{12}^\up > 0 &:& \varphi_\up^- = \frac{(2n+1)\pi}{2 \sqrt{2}}, \quad n \in \mathbb{Z} ~, \\
&& \epsilon_{k_{F1\up} + k_{F2\up}} \sim e^{i \theta_{\sigma+}},\;\;  \chi_{k_{F1\up} + k_{F2\up}} = 0; \label{w12+ve} \\
w_{12}^\up < 0 &:& \varphi_\up^- = \frac{2n\pi}{2 \sqrt{2}}, \quad n \in \mathbb{Z} ~, \\
&& \chi_{k_{F1\up} + k_{F2\up}} \sim e^{i \theta_{\sigma+}},\;\; \epsilon_{k_{F1\up} + k_{F2\up}} = 0 ~. \label{w12-ve}
\end{eqnarray}

Next, note that all spin-1 observables $S^+_Q$ become short-ranged since they all contain the wildly fluctuating field $\theta_\up^-$.  Schematically, the individual $f_\up$ become gapped because of their ``pairing''.  On the other hand, spin-2 observables contain pairs of $f_\up$ and can remain gapless.  Explicitly, after pinning the $\varphi_\up^-$, we have for the dominant correlations at $Q=0$ and $\pi$
\begin{eqnarray}
{\cal P}^{+}_{Q=0} &\sim& e^{-i 2\varphi_{\sigma+}} e^{\pm i \sqrt{2}\varphi_\dn^-} ~,\\
{\cal P}^{+,(1)}_{Q=\pi} &\sim& e^{-i 2\varphi_{\sigma+}}~.
\end{eqnarray}

The gaplessness of the $\varphi_{\sigma+}$ is required since $S^z_{\rm tot}$ is conserved and incommensurate with the lattice.  We can map the spin system to hard-core bosons,\cite{Hikihara08} and in the present case single boson excitations are gapped, while pair boson excitations are gapless and created by $e^{i 2\varphi_{\sigma+}} \dots$.  In the ``particle-hole'' sector, we have strong ``density'' or ``current'' correlations, Eq.~(\ref{w12+ve}) or (\ref{w12-ve}), at wavelengths that can be related to typical separations between boson pairs, and such $e^{i\theta_{\sigma+}}$ contribution is generally expected in a Luttinger liquid of pairs.  Thus, the resulting state has spin-nematic power law correlations as well as density or current power law correlations.  Which one is dominant depends on the scaling dimensions of $e^{i 2\varphi_{\sigma+}}$ versus $e^{i \theta_{\sigma+}}$.  The scaling dimensions would need to be calculated numerically since the $\varphi_{\sigma+}$ and $\varphi_\dn^-$ mix in general; we do not attempt such quantitative estimates here.

Having discussed observables controlled by the gapless $\sigma+$ part, let us finally mention that ${\cal B}^{(1)}_\pi$ and $\chi_\pi$ directly detect the gapless $\varphi_\dn^-$ field, cf.\ Eqs.~(\ref{B1pi1})-(\ref{chipi2b}).  In the phase discussed in this section they have the same power law decays.

\begin{table}
\begin{tabular}{|c|c|c|c|c|}
\hline
\multicolumn{5}{|c|}{Pinned $\varphi_\up^-$: Common power-law order for either sign of $w_{12}^\up$}
\\
\hline
\multicolumn{1}{|c|}{\multirow{2}{*}{$\epsilon_{\pm 2 k_{Fa\dn}}$}}&
\multicolumn{1}{|c|}{$\epsilon_{\pm (k_{F1\dn} + k_{F2\dn})}$;}&
\multicolumn{1}{|c|}{$\epsilon_{\pm (k_{F1\dn} - k_{F2\dn})}$;}&
\multicolumn{1}{|c|}{${\cal B}^{(1)}_{\pi}$;}&
\multicolumn{1}{|c|}{\multirow{2}{*}{${\cal P}^+_{\{Q\}}$}}
\\
\multicolumn{1}{|c|}{\multirow{2}{*}{}}&
\multicolumn{1}{|c|}{$\chi_{\pm (k_{F1\dn} + k_{F2\dn})}$}&
\multicolumn{1}{|c|}{$\chi_{\pm (k_{F1\dn} - k_{F2\dn})}$}&
\multicolumn{1}{|c|}{$\chi_{\pi}$}&
\multicolumn{1}{|c|}{\multirow{2}{*}{}}
\\ \hline\hline
\multicolumn{5}{|c|}{Distinct power law correlations} \\
\hline
\multicolumn{5}{|c|}{$w_{12}^\up > 0: \quad \epsilon_{\pm (k_{F1\up} + k_{F2\up})}$} \\
\multicolumn{5}{|c|}{$w_{12}^\up < 0: \quad \chi_{\pm (k_{F1\up} + k_{F2\up})}$}
\\ \hline
\end{tabular}
\caption{Summary of the main observables when $w^\up_{12}$ term is relevant and pins $\varphi_\up^-$. Critical wavevectors $Q$ for the magnon-pair creation operator are obtained by combining any of $q_\up = \{0, ~\pm(k_{F1\up} + k_{F2\up})\}$ with any of $q_\dn = \{0, ~\pm(k_{F1\dn} + k_{F2\dn}), ~\pm(k_{F1\dn} - k_{F2\dn})\}$, $Q = q_\up + q_\dn$; the most important ones are $Q=0$ and $\pi$.
}
\label{tab:up-relevant-comm}
\end{table}

We have considered the case when only $w^\up_{12}$ becomes relevant.  The case when only $w^\dn_{12}$ becomes relevant can be treated similarly by interchanging $\up$ and $\dn$.

\subsection{Phases when both $w^\up_{12}$ and $w^\dn_{12}$ are relevant} \label{bothw12relevant}
Let us now discuss the phases out of the SBM when both $w^\up_{12}$ and $w^\dn_{12}$ terms get relevant.  Once the couplings flow to large values, both variables $\varphi_\up^-$ and $\varphi_\dn^-$ will be pinned so as to minimize the energy.  There are four possible situations depending on the signs of the $w^\up_{12}$ and $w^\dn_{12}$.

In all cases, we find that the translational symmetry is broken by either a static order in ${\cal B}_\pi$ (corresponding to period-2 valence bond solid) or $\chi_\pi$ (corresponding to period-2 chirality order).  Coexisting with this, we have one gapless mode, namely the overall spin mode ``$\sigma+$'', which must remain gapless as long as the magnetization density is incommensurate with the lattice.  Similarly to the case with one relevant coupling, spin-1 observables are gapped.  Spin-2 observables are gapless, with the dominant contributions
\begin{equation}
{\cal P}_{Q=0}^+ \sim {\cal P}_{Q=\pi}^+ \sim e^{-i 2\varphi_{\sigma+}} ~.
\end{equation}
(Note that the original wavevectors $Q=0$ and $\pi$ are not distinguishable once we have the period-2 static orders.)  Together with such spin-nematic observables, we also have spin-0 observables of the $\epsilon$- or $\chi$-type depending on the pinning details, with the wavevectors $\pm (k_{F1\alpha} + k_{F2\alpha})$ which satisfy $k_{F1\up} + k_{F2\up} = -(k_{F1\dn} + k_{F2\dn}) - \pi$.

Below, we consider four different pinning situations in more details.  The main features in each case are summarized in Table~\ref{tab:B-relevant}.

\begin{table}
\begin{tabular}{|c|c|c|c|c|}
\hline
\multicolumn{1}{|c|}{$w^\up_{12}$} &
\multicolumn{1}{|c|}{$w^\dn_{12}$} &
\multicolumn{1}{|c|}{Static Order} &
\multicolumn{2}{|c|}{Power-Law Correlations}
\\
\hline
\multicolumn{1}{|c|}{+} &
\multicolumn{1}{|c|}{+} &
\multicolumn{1}{|c|}{${\cal B}^{(1)}_\pi$} &
\multicolumn{1}{|c|}{$\epsilon_{\pm(k_{F1\alpha} + k_{F2\alpha})}$}&
 \multicolumn{1}{|c|}{${\cal P}^+_{\{Q\}}$}
\\
\hline
\multicolumn{1}{|c|}{-} &
\multicolumn{1}{|c|}{-} &
\multicolumn{1}{|c|}{${\cal B}^{(1)}_\pi$} &
\multicolumn{1}{|c|}{$\chi_{\pm(k_{F1\alpha} + k_{F2\alpha})}$}&
\multicolumn{1}{|c|}{${\cal P}^+_{\{Q\}}$}
\\
\hline
\multicolumn{1}{|c|}{\multirow{2}{*}{+}} &
\multicolumn{1}{|c|}{\multirow{2}{*}{-}} &
\multicolumn{1}{|c|}{\multirow{2}{*}{$\chi_\pi$}} &
\multicolumn{1}{|c|}{$\epsilon_{\pm (k_{F1\up} + k_{F2\up})}$;}&
\multicolumn{1}{|c|}{\multirow{2}{*}{${\cal P}^+_{\{Q\}}$}}
\\
\multicolumn{1}{|c|}{\multirow{2}{*}{}} &
\multicolumn{1}{|c|}{\multirow{2}{*}{}} &
\multicolumn{1}{|c|}{\multirow{2}{*}{}} &
\multicolumn{1}{|c|}{$\chi_{\pm(k_{F1\dn} + k_{F2\dn})}$}&
\multicolumn{1}{|c|}{\multirow{2}{*}{}}
\\
\hline
\multicolumn{1}{|c|}{\multirow{2}{*}{-}} &
\multicolumn{1}{|c|}{\multirow{2}{*}{+}} &
\multicolumn{1}{|c|}{\multirow{2}{*}{$\chi_{\pi}$}} &
\multicolumn{1}{|c|}{$\epsilon_{\pm(k_{F1\dn} + k_{F2\dn})}$;}&
\multicolumn{1}{|c|}{\multirow{2}{*}{${\cal P}^+_{\{Q\}}$}}
\\
\multicolumn{1}{|c|}{\multirow{2}{*}{}} &
\multicolumn{1}{|c|}{\multirow{2}{*}{}} &
\multicolumn{1}{|c|}{\multirow{2}{*}{}} &
\multicolumn{1}{|c|}{$\chi_{\pm(k_{F1\up} + k_{F2\up})}$}&
\multicolumn{1}{|c|}{\multirow{2}{*}{}}
\\
\hline
\end{tabular}
\caption{Summary of the cases when both $w^\up_{12}$ and $w^\dn_{12}$ terms are relevant.  For $w^\up_{12} w^\dn_{12} > 0$ we have period-2 VBS order, while for $w^\up_{12} w^\dn_{12} < 0$ we have period-2 chirality order.  In all cases, coexisting with such static order, we have power law correlations in the spin-2 (magnon pair) observable ${\cal P}^+$ and in the specific $\epsilon/\chi$ observables.
}
\label{tab:B-relevant}
\end{table}

\subsubsection{$w^\up_{12} > 0$, $w^\dn_{12} > 0$}\label{wgg0}

\begin{figure}[t]
  \includegraphics[width=\columnwidth]{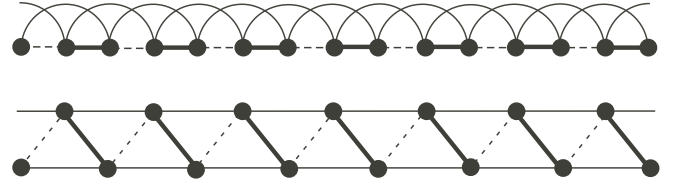}
  \caption{Picture of the Valence Bond Solid order when ${\cal B}^{(1)}_{\pi}$ gains an expectation value.  Top: 1d chain view.  Bottom: the same in 2-leg ladder view.  Coexisting with the static order, we also have spin-nematic power law correlations and power law in either $\epsilon$ or $\chi$ channels (these properties are not depicted in any way).}
  \label{fig:vbs}
\end{figure}

The pinning conditions for fields $\varphi_\up^-$ and $\varphi_\dn^-$ are
\begin{eqnarray}
\varphi_\up^- = \frac{(2n+1)\pi}{2 \sqrt{2}}, \quad
\varphi_\dn^- = \frac{(2m+1)\pi}{2 \sqrt{2}}, \quad n, m\in\mathbb{Z} ~.
\end{eqnarray}
In this case, ${\cal B}^{(1)}_\pi$ obtains an expectation value while $\chi_\pi=0$.  Thus we expect to see period-2 VBS order as illustrated in Fig.~\ref{fig:vbs}.  We also have power law correlations in
\begin{equation}
\epsilon_{k_{F1\up} + k_{F2\up}} \sim \epsilon_{-k_{F1\dn} - k_{F2\dn}} \sim e^{i\theta_{\sigma+}} ~,
\end{equation}
while $\chi_{k_{F1\alpha} + k_{F2\alpha}} = 0$.  Note that because of the relation Eq.~(\ref{B1pi1}) [in the sense that $i(\epsilon_{k_{F1\up} + k_{F2\up}} \epsilon_{k_{F1\dn} + k_{F2\dn}} - \Hc)$ has the same symmetry properties as ${\cal B}^{(1)}_\pi$], once the system develops static order in ${\cal B}^{(1)}_\pi$, the $\epsilon_{k_{F1\up} + k_{F2\up}}$ and $\epsilon_{-k_{F1\dn} - k_{F2\dn}} = \epsilon_{k_{F1\dn} + k_{F2\dn}}^\dagger$ are no longer independent.  Appropriately, the wavevectors $k_{F1\up} + k_{F2\up}$ and $-k_{F1\dn} - k_{F2\dn}$ differ by $\pi$ and also become connected.

\subsubsection{$w^\up_{12} < 0$, $w^\dn_{12} < 0$}

Here, the pinning conditions are
\begin{eqnarray}
\varphi_\up^- = \frac{2n\pi}{2 \sqrt{2}}, \quad
\varphi_\dn^- = \frac{2m\pi}{2 \sqrt{2}}, \quad n, m\in\mathbb{Z} ~.
\end{eqnarray}
Again, ${\cal B}^{(1)}_\pi$ obtains an expectation value while $\chi_\pi=0$.  However, here we have power law correlations in
\begin{equation}
\chi_{k_{F1\up} + k_{F2\up}} \sim \chi_{-k_{F1\dn} - k_{F2\dn}} \sim e^{i\theta_{\sigma+}}
\end{equation}
while $\epsilon_{k_{F1\alpha} + k_{F2\alpha}} = 0$.  Similar to the discussion in the preceding case and using relation Eq.~(\ref{B1pi2}), $\chi_{k_{F1\up} + k_{F2\up}}$ and $\chi_{-k_{F1\dn} - k_{F2\dn}}$ are not independent observables in the presence of the static order in ${\cal B}^{(1)}_\pi$.

\subsubsection{$w^\up_{12} > 0$, $w^\dn_{12} < 0$}

\begin{figure}[t]
  \includegraphics[width=\columnwidth]{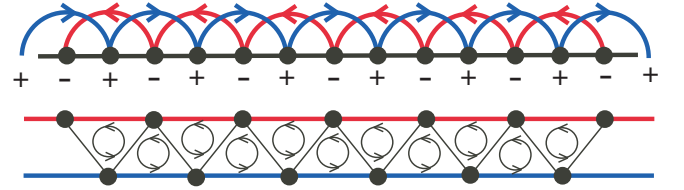}
  \caption{
Picture of the static period-2 order in spin chirality when $\chi_\pi$ gains an expectation value.  Since ${\cal J}^{(2)}_\pi \sim \chi_\pi$, we have static staggered second-neighbor bond currents in the chain view (top figure).  In the ladder view (bottom figure), we have oppositely oriented spin currents flowing on the two legs.  Coexisting with the static order, we also have spin-nematic power law correlations and power laws in $\epsilon/\chi$ channels (these properties are not depicted in any way).}
  \label{chiralityphase}
\end{figure}

In this case, the pinning conditions are
\begin{eqnarray}
\varphi_\up^- = \frac{(2n+1)\pi}{2 \sqrt{2}}, \quad
\varphi_\dn^- = \frac{2m\pi}{2 \sqrt{2}}, \quad n, m\in\mathbb{Z} ~.
\end{eqnarray}
In this phase, $\chi_\pi$ obtains an expectation value while ${\cal B}^{(1)}_\pi = 0$.  Thus we expect to see period-2 chirality order as illustrated in Fig.~\ref{chiralityphase}.  We also have power law correlations in
\begin{equation}
 \epsilon_{k_{F1\up} + k_{F2\up}} \sim \chi_{-k_{F1\dn} - k_{F2\dn}} \sim e^{i \theta_{\sigma+}} ~,
\end{equation}
while $\chi_{k_{F1\up} + k_{F2\up}} = \epsilon_{-k_{F1\dn} - k_{F2\dn}} = 0$.  By using Eq.~(\ref{chipi1}), we can understand the equivalence of the two observables $\epsilon_{k_{F1\up} + k_{F2\up}}$ and $\chi_{-k_{F1\dn} - k_{F2\dn}}$ once there is the static order in $\chi_\pi$.

\subsubsection{$w^\up_{12} < 0$, $w^\dn_{12} > 0$}

In this case, the pinning conditions are
\begin{eqnarray}
\varphi_\up^- = \frac{2n\pi}{2 \sqrt{2}}, \quad
\varphi_\dn^- = \frac{(2m+1)\pi}{2 \sqrt{2}}, \quad n, m\in\mathbb{Z} ~.
\end{eqnarray}
$\chi_\pi$ obtains an expectation value while ${\cal B}^{(1)}_\pi = 0$.  We also have power law correlations in
\begin{equation}
 \chi_{k_{F1\up} + k_{F2\up}} \sim \epsilon_{-k_{F1\dn} - k_{F2\dn}} \sim e^{i \theta_{\sigma+}} ~,
\end{equation}
while $\epsilon_{k_{F1\up} + k_{F2\up}} = \chi_{-k_{F1\dn} - k_{F2\dn}} = 0$.  The two observables $\chi_{k_{F1\up} + k_{F2\up}}$ and $\epsilon_{-k_{F1\dn} - k_{F2\dn}}$ become related because of Eq.~(\ref{chipi2}) and the static order in $\chi_\pi$.

This completes our discussion of the phases out of the SBM.  We cannot tell which of the different cases are more likely in particular microscopic models.  Also, the power law correlation exponents depend on the unknown Luttinger parameter $g_{\sigma+}$ of the ``$\sigma+$'' field, and we cannot tell whether spin-2 or spin-0 observables dominate (their scaling dimensions are $1/g_{\sigma+}$ and $g_{\sigma+}/4$ respectively).  However, we have developed a qualitative understanding of the phases and observables needed to identify them, which we hope will be useful in numerical studies of models realizing the SBM phase.

\section{Discussion}

In this paper, we studied instabilities of the 2-leg SBM under the Zeeman magnetic field.  The instabilities are driven by the $w_{12}^\alpha$ interactions, Eq.~(\ref{w12}), and we analyzed possible outcomes using Bosonization.  In all cases, we found a gap to spin-1 excitations, while spin-nematic (two-magnon) correlations are power law.  Loosely speaking, this appears because of some pairing of spinons, while the precise characterization is obtained by analyzing all physical observables.

Here we want to discuss consequences if such spinon pairing were to occur in a 2D spin liquid under the Zeeman field.  At present, we do not have any energetics justification under which circumstances this may happen and whether this applies to the candidate spin liquid materials.  However, the resulting states are quite interesting on their own and perhaps such phases may occur in some other 2D systems (several papers\cite{GalKim07, Grover09} considered mechanisms for spinon pairing in zero field).

First of all, the analog of the stable SBM phase in Sec.~\ref{strong-coupling:Zeeman} has gapless Fermi surfaces for both $\up$ and $\dn$ spinon species, with somewhat different $k_{F\up}$ and $k_{F\dn}$.  In the organic \ET\ and \dmit\ materials, we estimate $(n_\up - n_\dn)/(n_\up + n_\dn) < 0.02$ under laboratory fields, so the difference between the two Fermi surfaces is small.  In mean field, the spin correlations are
\begin{eqnarray}
  \left\la S^+({\bf r}) S^-({\bf 0}) \right\ra_{\rm mf} &\sim& -\frac{\cos[({\bf k}_{F\up} + {\bf k}_{F\dn}) \cdot {\bf r} + \frac{\pi}{2}]}{|{\bf r}|^3} \\
  &&- \frac{\cos[({\bf k}_{F\up} - {\bf k}_{F\dn}) \cdot {\bf r}]}{|{\bf r}|^3} ~, \\
  \left\la \delta S^z({\bf r}) \delta S^z({\bf 0}) \right\ra_{\rm mf} &\sim& -\sum_{\alpha=\up,\dn} \frac{1 + \cos[2 {\bf k}_{F\alpha} \cdot {\bf r} + \frac{\pi}{2}]}{|{\bf r}|^3} ~, \hspace{0.5cm}
\end{eqnarray}
while gauge fluctuations are expected to enhance the ${\bf k}_{F\up} + {\bf k}_{F\dn}$ and $2 {\bf k}_{F\alpha}$ parts,\cite{Altshuler94} similarly to the ladder case.\cite{Sheng09}

Next, we want to discuss the analog of the situation in Sec.~\ref{onew12relevant}, where there is pairing in one spinon species (say, $f_\up$) and no pairing in the other species.  Note that the pairing must be odd-wave since it is within one fermion type.  We will not consider any energetics selection of the pairing and just mention possibilities like $p$-wave ($p_x + i p_y$) or $f$-wave that can be nicely placed on the triangular lattice.  

The properties of the resulting phase are as follows.  The $f_\dn$ species are gapless with Fermi surface, so we expect metal-like specific heat $C = \gamma T$; note that this is the full result since the gauge field is Higgsed out by the $f_\up$ pairing.  We also expect constant spin susceptibility at $T \to 0$ since both $f_\up$ and $f_\dn$ systems are compressible, the former due to the pair-condensate and the latter by virtue of finite density of states at the Fermi level.  Because of the $f_\dn$ Fermi surface, we expect $\la S^z({\bf r}) S^z ({\bf 0})\ra$ to show $2k_{F\dn}$ oscillations with $1/r^3$ power law.  On the other hand, $\la S^+({\bf r}) S^-({\bf 0}) \ra$ will show either a full gap if the $f_\up$ pairing is fully gapped as in the case of $p_x + i p_y$ pairing, or a pseudogap if the $f_\up$ pairing has gapless parts as in the case of $f$-wave pairing.  Note that this does not contradict the finite susceptibility since the $f_\up$-pair condensate can readily accommodate $\Delta N_\up = \pm 2$ changes.  Related to this, spin-nematic correlations are gapless and show $1/r^3$ power law at zero wavevector (in the mean field calculation).  Interestingly, the gap or pseudo-gap to spin-1 operators would have consequences for NMR experiments done with $^1$H or $^{13}$C that are both spin-$\frac{1}{2}$ nuclei and relax only by spin-1 excitations.  From such measurements, this phase might appear gapped, but it actually has a gapless Fermi surface of one species.  (In the context of 1D models exhibiting spin-nematic phases, consequences for the NMR relaxation rate were discussed in detail e.g.\ in Ref.~\onlinecite{Sato09}.)

Finally, let us consider the analog of the situation in Sec.~\ref{bothw12relevant}, where both $f_\up$ and $f_\dn$ become paired, with possibly different pairing $\Delta^\up_{{\bf r} {\bf r}'}$, $\Delta^\dn_{{\bf r} {\bf r}'}$.  In this case, $S^z$ and $S^+$ correlations are both gapped (or pseudo-gapped), while spin-nematic correlation shows long-range order. Specifically, in the mean field, 
\begin{eqnarray}
\left\la S^+({\bf r}) S^+({\bf r}') \right\ra_{\rm mf} = \Delta^{\up *}_{{\bf r} {\bf r}'} \Delta^\dn_{{\bf r} {\bf r}'} ~.
\end{eqnarray}
Note that this nematic order resides on the bonds of the lattice and details depend on the $\Delta^\up$ and $\Delta^\dn$.  For example, if we take $\Delta^\up$ and $\Delta^\dn$ to have the same pattern, this will give ferro-nematic state.  Curiously, if we take $\Delta^\up \sim p_x + i p_y$ and $\Delta^\dn \sim p_x - i p_y$, we get $q=0$ antiferromagnetic nematic order on the Kagome lattice formed by the bonds of the triangular lattice. We emphasize that we have not discussed any energetics that may be selecting among such states.  Whether something like this can appear in realistic models on the triangular lattice is an interesting open question.

\section{Acknowledgement}

We would like to thank C.-C.~Chen, M.~P.~A.~Fisher and P.~A.~Lee for useful discussions. This research is supported by the National Science Foundation through grant DMR-0907145 and by the A.~P.~Sloan Foundation.

\bibliography{biblio4zigzag}

\end{document}